\documentclass[aps,prb,preprint,superscriptaddress]{revtex4}
\usepackage{amsfonts}
\usepackage{amsmath}
\usepackage{amssymb}
\usepackage{multirow}
\usepackage{graphicx}
\usepackage{dcolumn}
\usepackage{bm}
\usepackage{titlesec}
\titleformat*{\section}{\flushleft \bf \large}
\titleformat*{\subsection}{\flushleft \bf}
\titleformat*{\subsubsection}{\flushleft}
\begin{document}


\title{Electron correlation and spin-orbit coupling effects in US$_3$ and USe$_3$}

\author{Yu Yang}
\affiliation{LCP, Institute of Applied Physics and Computational
Mathematics, P.O. Box 8009, Beijing 100088, People's Republic of
China}
\author{Ping Zhang}
\thanks{Corresponding author. zhang\_ping@iapcm.ac.cn}
\affiliation{LCP, Institute of Applied Physics and Computational
Mathematics, P.O. Box 8009, Beijing 100088, People's Republic of
China}%
\affiliation{Center for Applied Physics and Technology,
Peking University, Beijing 100871, People's Republic of China}%

\date{\today}

\begin{abstract}

A systematic density functional theory (DFT) +$U$ study is conducted
to investigate the electron correlation and spin-orbit coupling
(SOC) effects in US$_3$ and USe$_3$. Our calculations reveal that
inclusion of the $U$ term is essential to get energy band gaps for
them, indicating the strong correlation effects for uranium 5$f$
electrons. Taking consideration of the SOC effect results in small
reduction on the electronic band gaps of US$_3$ and USe$_3$, but
largely changes the energy band shapes around the Fermi energy. As a
result, US$_3$ has a direct band gap while USe$_3$ has an indirect
one. Our calculations predict that both US$_3$ and USe$_3$ are
antiferromagnetic insulators, in agreement with corresponding
experimental results. Based on our DFT+$U$ calculations, we
systematically present the ground-state electronic, mechanical and
Raman properties for US$_3$ and USe$_3$.

\end{abstract}

\maketitle

\section{Introduction}

Actinide based materials possess interesting physical behaviors due
to the existence of strongly-correlated 5$f$ electrons and have
attracted extensive attentions
\cite{Savrasov01,Albers01,Hecker04,Moore06,Prodan06,Prodan07}.
Different from actinide oxides which have been widely studied to
reveal the detailed configurations and correlation effects of the
5$f$ electrons
\cite{Prodan06,Prodan07,Dudarev97,Dudarev98,Dudarev00,Sun08JCP,Sun08CPB,Shi10,WangBT10,Zhang10,Kern99,Petit03,Petit10,Wilkins06,Yin08,Jomard08,Andersson09,Santini09,Sanati11,Dorado09,Dorado10,Nakamura10,Meredig10,Geng10,Yasuoka12,Manley12},
none of the actinide chalcogenides has ever received comparable
concerns. Here we take US$_3$ and USe$_3$ as two representatives to
study the electron correlation and spin-orbit coupling (SOC) effects
of 5$f$ electrons in actinide chalcogenides. The other reason for us
to investigate their electronic structures is that they employ the
layered MX$_3$ structure (with M to be a metal element, and X to be
S, Se, or Te), which can be used as models for studying electronic
behaviors in 1-dimensional (1D) systems. The MX$_3$ structure
belongs to the space group of P2$_1$/$m$, with its atomic
organizations depicted in Fig. 1(a). In the monoclinic lattice, the
top two chalcogen (S , Se or Te) atoms form a tightly bound pair
suggesting that the MX$_3$ compounds may be regarded as MX(X$_2$)
\cite{Nouvel87}. The top X-X pair together with an underneath M and
a more underneath X atom form a triangular unit, which repeats along
the $\vec{b}$ lattice direction forming a 1D chain, and two such
triangle units form a unit cell of the MX$_3$ compounds in the
($\vec{a}$,$\vec{c}$) plane. In this way, the MX$_3$ compounds are
always considered as 1D materials. For example, the metallic
ZrTe$_3$ and NbSe$_3$ exhibiting charge-density-wave transitions
have been thoroughly investigated in angle-resolved photoelectron
spectroscopy (ARPES) and optical experiments
\cite{Schafer03,Perucchi041,Perucchi042,Yokoya05,Pacile07}. In the
present paper, by contrast, we address the electronic structures of
US$_3$ and USe$_3$, two semiconducting members of the MX$_3$ family.

In studies of actinide based materials, one has to be careful for
the electron correlation and SOC effects of the actinide 5$f$
electrons. As an example, for actinide dioxides conventional density
functional theory (DFT) schemes that apply the local density
approximation (LDA) or the generalized gradient approximation (GGA)
underestimate the strong on-site Coulomb repulsion of the 5$f$
electrons and consequently fail to capture the insulating properties
\cite{Dudarev97,Zhang10}. Several approaches, the LDA/GGA+$U$, the
hybrid density functional of (Heyd, Scuseria, and Enzerhof) HSE, the
self-interaction corrected local spin-density (SIC-LSD), and the
Dynamical Mean-Field Theory (DMFT), have been developed to correct
the pure LDA/GGA failures in calculations of actinide materials.
Among them the effective modification of pure DFT by LDA/GGA+$U$
formalisms has been widely used in theoretical studies of UO$_2$
\cite{Dudarev97,Dudarev00,Sanati11} and PuO$_2$
\cite{Sun08JCP,Sun08CPB,Jomard08,Shi10}. The obtained structural
parameters as well as the electronic structure and phonon dispersion
curves \cite{Zhang10,Manley12} accord well with experiments. In our
present work, the GGA+$U$ schemes due to Dudarev {\it et al.}
\cite{Dudarev97,Dudarev98,Dudarev00} are employed to study the
electron correlation and spin-orbit coupling effects in US$_3$ and
USe$_3$, as well as the two materials' electronic, mechanical, and
Raman properties.

The rest of the paper is organized as follows. In Sec. II the
computational method is briefly described. In Sec. III we present
the results of the physical properties of US$_3$ and USe$_3$, and
discuss the electron correlation and SOC effects of the uranium 5$f$
electrons. Finally in Sec. IV, we close our paper with a summary of
our main results.

\section{Calculation method}

Our total-energy calculations are carried out by employing the
plane-wave basis pseudopotential method as implemented in Vienna ab
initio simulation package (VASP) \cite{VASP}. The exchange and
correlation effects are described with the GGA approximation in the
Perdew-Burke-Ernzerhof (PBE) form \cite{PBE1,PBE2}. The projected
augmented wave (PAW) method of Bl\"{o}chl \cite{PAW} is employed
with the frozen-core approximation. Electron wave function is
expanded in plane waves up to a cutoff energy of 400 eV and all
atoms are fully relaxed until the Hellmann-Feynman forces on them
are less than 0.01 eV/\AA. A 7$\times$9$\times$1 Monkhorst-Pack
\cite{MPkpoints} k-point mesh is employed for integration over the
Brillouin zone of US$_3$ and USe$_3$. The uranium
6$s^2$6$p^6$5$f^3$6$d^1$7$s^2$, sulphur 3$s^2$3$p^6$, and selenium
4$s^2$4$p^6$ electrons are treated as valence electrons.
Noncollinear calculations are used when considering the spin-orbital
coupling effects. The strong on-site Coulomb repulsions among the
localized uranium 5$f$ electrons are described by using the
formalism formulated by Dudarev et al.
\cite{Dudarev97,Dudarev98,Dudarev00}. In this scheme, the total GGA
energy functional is of the form
\begin{equation}
    E_{{\rm GGA}+U}=E_{\rm GGA}+\frac{U-J}{2}\sum_{\sigma} [{\rm Tr}\rho^\sigma-{\rm Tr}(\rho^\sigma\rho^\sigma)],
\end{equation}
where $\rho^\sigma$ is the density matrix of $f$ states with spin
$\sigma$, while $U$ and $J$ are the spherically averaged screened
Coulomb energy and the exchange energy, respectively.

In this paper, the Coulomb $U$ is treated as a variable, while the
exchange energy is set to be a constant $J$=0.51 eV. This value of
$J$ is in the ball park of the commonly accepted one for uranium
compounds \cite{Sanati11} and close to the theoretically predicted
value of 0.54 eV in UO$_2$ \cite{Kotani92}. Since only the
difference between $U$ and $J$ is significant, we will henceforth
label them as one single parameter $U_{\rm eff}$=$U$-$J$, while
keeping in mind that the nonzero $J$ has been used during
calculations.

The phonon frequencies at the Gamma point for US$_3$ and USe$_3$ are
calculated by using the density functional perturbation theory
(DFPT). And their Raman-active frequencies are obtained through
symmetry analysis on the corresponding vibration modes. Before DFPT
calculations, the lattice constants and atomic positions of US$_3$
and USe$_3$ are further optimized using a denser $k$-point mesh of
9$\times$13$\times$3, and a finer force convergence criteria of
0.001 eV/\AA. A 2$\times$2$\times$1 supercell is subsequently used
for DFPT calculations. Since both US$_3$ and USe$_3$ are stacked
along the $\vec{c}$ direction through van der Waals interactions,
and with relatively large lattice constants, we do not extend the
supercell along the $\vec{c}$ direction. During DFPT calculations on
the 2$\times$2$\times$1 supercell, the $k$-point mesh is set to be
5$\times$7$\times$5, and the energy convergence criteria is set to
be 1e-6 eV.

\section{Results and discussion}

Experimentally, the magnetic orderings of uranium trichalcogenides
were studied ever since 1961
\cite{Nouvel87,Trzebiatowski61,Suski76,Baran83,Noel86}. The
relatively large atomic distances between neighboring U atoms in
US$_3$ and USe$_3$ suggest that magnetic orderings may occur in
these compounds through super-exchange interactions of uranium 5$f$
electrons via the S or Se ions \cite{Nouvel87}. And different
measurements indicate that both US$_3$ and USe$_3$ crystals undergo
AFM transitions at very low temperatures, with the magnetic moments
of the two uranium atoms within each unit cell opposite to each
other \cite{Nouvel87,Baran83,Noel86}. In the present GGA+$U$ study,
we consider the nonmagnetic (NM), ferromagnetic (FM), and
antiferromagnetic (AFM) phases for each choice of the $U_{\rm eff}$
value and then determine the lowest-energy state by a subsequent
total-energy comparison of these three phases. Our calculated
electronic energies for different magnetic phases of US$_3$ and
USe$_3$ are all listed in Table I. One can see that within GGA
formalism or the GGA+$U$ formalism with a too small $U_{\rm eff}$
value, the FM state is more stable for both US$_3$ and USe$_3$, in
contradiction with experimental results. Therefore, the electron
correlation effect has to be accounted for to correctly describe
ground-state US$_3$ and USe$_3$. In the discussions that follow, we
will confine our reports to the AFM solutions for US$_3$ and
USe$_3$.

The experimentally determined lattice parameters of US$_3$ and
USe$_3$ are ($a$=5.37 \AA, $b$=3.96 \AA, $c$=9.94 \AA) and ($a$=5.65
\AA, $b$=4.06 \AA, $c$=10.47 \AA) respectively
\cite{Nouvel87,Gronvold68,Salem84}. Our calculated lattice constants
with different $U_{\rm eff}$ values are summarized in Table II,
together with the experimental results. We can see that different
from what we found for actinide dioxides \cite{Zhang10,Wang10}, the
lattice constants of actinide trichalcogenides do not change
monotonically with the value of the $U_{\rm eff}$ parameter.
Besides, the lattice constants obtained in GGA calculations are
obviously too small in comparison with corresponding experimental
results, for both US$_3$ and USe$_3$. For US$_3$, the value of
$U_{\rm eff}$ has to be as large as 6 eV to get reasonable lattice
constants compared with experimental results. Differently, a $U_{\rm
eff}$ value of 4 eV is enough to get reasonable lattice constants
for USe$_3$, and changing $U_{\rm eff}$ from 4 to 6 eV has
negligible effects on the obtained lattice constants.

Besides of the prominent changes in the atomic structure parameters,
the most dramatic improvement brought by the GGA+$U$ formalism when
compared to the GGA results is in the description of electronic
structure properties. For this, we have investigated the band
structures in AFM phases for US$_3$ and USe$_3$ aiming at revealing
the fundamental influences of considering the on-site Coulomb
interaction. Figures 2(a) and 2(b) show the obtained local density
of states (LDOS) for the S1, S3, and U atoms in US$_3$, and Se1,
Se3, and U atoms in USe$_3$ respectively. Without accounting for the
on-site Coulomb repulsion, one can see that the GGA calculations
predict an incorrect metallic ground state by nonzero occupation of
uranium electrons at the Fermi energy ($E_{\rm f}$). When switching
on the $U_{\rm eff}$ parameter, as shown both in Figs. 2(a) and
2(b), the uranium electronic states in US$_3$ and USe$_3$ begin to
split at $E_{\rm f}$ and tend to form two peaks with the gap of
$\Delta E$. The amplitude of this gap increases with increasing
$U_{\rm eff}$. Previous electric resistivity measurements pointed
out that USe$_3$ did not conduct at room temperatures
\cite{Shlyk95}, thus a band gap at $E_{\rm f}$ should be contained
in its electronic structure. We can see from Fig. 2(b) that an
energy band gap appears only when $U_{\rm eff}$ is larger than or
equal to 4 eV for USe$_3$. For US$_3$, the value of $U_{\rm eff}$
has to be enlarged to 6 eV to open the band gap at the Fermi energy.
These results also imply that the electron correlation strength
might be different in US$_3$ and USe$_3$. From the last figure in
Figs. 2(a) and 2(b), we can see that the sulphur (selenium) and
uranium electronic states overlap with each other and contribute
equally to the valence band maximum (VBM) of US$_3$ (USe$_3$), while
the conduction band minimum (CBM) is composed of uranium electronic
states. The orbital mixing between sulphur (selenium) and uranium
electronic states below the Fermi energy indicates that there are
covalent interactions between the sulphur (selenium) and uranium
atoms in US$_3$ (USe$_3$).

To further analyze the orbital-resolved electronic structures, we
calculate the projected density of states (PDOS) for the X-$np$
($n$=3 (X=S) or 4 (X=Se)), U-5$f$ and U-6$p$ electronic states in
UX$_3$. Figures 3(a)-3(d) show the obtained PDOS for US$_3$ in
different calculations, while the corresponding PDOS results for
USe$_3$ are shown in Figs. 3(e)-3(h). The value of $U_{\rm eff}$ is
chosen to be 6 eV in all GGA+$U$ and GGA+$U$+SOC calculations. One
can see that without considering the electron correlation effect,
US$_3$ and USe$_3$ are both wrongly predicted as metallic materials,
which contradicts with experimental results. After adding the $U$
parameter to describe the strong on-site energy of U-5$f$ electrons,
the energy gaps are opened for US$_3$ and USe$_3$. We can see from
Figs. 3(d) and 3(h) that further considering SOC has little
influences on the electronic states around the Fermi energy. For
both US$_3$ and USe$_3$, the SOC effect lies in the deep energy
level, causing an energy splitting of the U-6$p$ states. From the
PDOS results, we can also see clear difference between the X1- and
X3-$np$ (n=3 (X=S) or 4(X=Se)) electronic states. The PDOS for the
electronic states of the X2 atom is very similar to that of the X1
atom and thus is not shown here. The different electronic state
distribution of the X1(X2) and X3 atoms proves the theory of
considering UX$_3$ as UX(X$_2$), and indicates the strong covalent
bondings between the X1 and X2 atoms.

To analyze more carefully the SOC effects in US$_3$ and USe$_3$, we
recalculate the electronic structures of them with higher
resolutions, with the reduced Gaussian smearing width of only 0.02
eV and a PDOS resolution of 0.01 eV. Figures 4(a) and 4(b) show the
obtained PDOS results for uranium 5$f$ electrons of US$_3$ and
USe$_3$, by using the GGA method, while Figs. 5(a) and 5(b) show the
obtained energy band structures of US$_3$ and USe$_3$ by using the
GGA+$U$ method, respectively. All the GGA+SOC and GGA+$U$+SOC
calculations are noncollinear, with both SOC and orbital
polarizations considered. And the results before and after
considering the SOC effects are shown in solid and dotted lines
respectively. From the PDOS results obtained by using the GGA
method, we can see that pure inclusion of SOC to the GGA method does
not lead to energy band gaps for US$_3$ and USe$_3$. This discovery
is different from the situation of CoO, where SOC can solely open a
band gap \cite{Norman90}.

From the band structures of US$_3$ shown in Fig. 5(a), we can see
that both the lowest unoccupied conduction and highest occupied
valence bands distribute along the $\Gamma$-Z direction. For the
lowest conduction band, the SOC effect causes an energy downshift of
0.05 eV along the $\Gamma$-Z direction. Contrarily for its highest
valence band, the SOC effect causes an energy upshift of 0.03 eV at
the $\Gamma$ point and an upshift of 0.07 eV at the Z point. As a
result, the energy band gap of US$_3$ changes from indirect
($\Gamma\rightarrow$Z) into direct type (Z$\rightarrow$Z) after
considering the SOC effect, and the gap value reduces from 1.59 to
be 1.48 eV. Different from US$_3$, the lowest conduction and highest
valence bands of USe$_3$ distributes along the K$_1$-B and
B-$\Gamma$-Y directions respectively. For the lowest conduction band
along the K$_1$-B direction, the SOC effect causes an energy
downshift of 0.04 eV at the K$_1$ point and a downshift of 0.02 eV
at the B point. Correspondingly the CBM of USe$_3$ moves from B to
the K$_1$ point. However, since the VBM of USe$_3$ is at some point
along the B-$\Gamma$-Y direction instead of at any high-symmetry $k$
points, the SOC effect does not change the indirect band gap type.
Overall the band gap of USe$_3$ changes from 1.30 to be 1.23 eV
after considering the SOC effect. Our studies reveal that the energy
band gap of USe$_3$ is smaller than that of US$_3$.

After systematically presenting the electronic structure results for
US$_3$ and USe$_3$, we now turn to their mechanical properties. The
elastic constants are obtained by solving the eigenvalues of their
Hessian matrix, which is calculated based on the Hooke's law and
small position changes on independent S, Se and U atoms. The
obtained elastic constants of US$_3$ and USe$_3$ in different
calculations by using the GGA and GGA+$U$ methods are both listed in
Table III. We can see that the elastic constants calculated by using
the GGA formalism for US$_3$ and USe$_3$ do not satisfy the
stability criteria \cite{Nye85,Wu07} for monoclinic structures that
the $C_{\rm 11}$, $C_{\rm 22}$, $C_{\rm 33}$, $C_{\rm 44}$, $C_{\rm
55}$, $C_{\rm 66}$ should be all positive. This result further
proves that conventional GGA formalism fails for describing the
uranium trichalcogenides. Contrarily, the calculated elastic
constants by using the GGA+$U$ method for both US$_3$ and USe$_3$
satisfy the above stability criteria, as well as the other six
criteria for monoclinic structures \cite{Nye85,Wu07}, proving the
stable existence of US$_3$ and USe$_3$, and their strong electron
correlation effects.

Based on the elastic constants of US$_3$ and USe$_3$, we further
calculate the Voigt and Reuss bounds on their bulk ($B_{\rm V}$,
$B_{\rm R}$) and shear moduli ($G_{\rm V}$, $G_{\rm R}$)
\cite{Supple}. And the bulk and shear moduli of US$_3$ and USe$_3$
can be estimated by $B$=($B_{\rm V}$+$B_{\rm R}$)/2, and
$G$=($G_{\rm V}$+$G_{\rm R}$)/2. And the Poisson's ratios can be
calculated by $\nu$=(3$B$-2$G$)/(6$B$+2$G$). The obtained mechanical
properties for US$_3$ and USe$_3$ are all listed in Table IV. We can
see that the Voigt and Beuss bounds obviously differ from each other
for bulk and shear moduli. This result reflects the structural
anisotropy of US$_3$ and USe$_3$. Besides, the bulk and shear moduli
are both found to be larger for US$_3$ than for USe$_3$. The
Poisson's ratio is calculated to be 0.19 and 0.22 for US$_3$ and
USe$_3$ respectively.

Until now, there are no experimental reports on the mechanical
properties of US$_3$ or USe$_3$. Hence our theoretical values can be
used as references for further investigations or industrial
applications of them. At another side, the Raman properties of the
MX$_3$ group materials have been systematically studied for a long
time, and some of the characteristic peaks of US$_3$ and USe$_3$
were successfully observed in experiments \cite{Nouvel87}.
Therefore, here we further calculate the Raman properties of US$_3$
and USe$_3$. According to group theory and the symmetry of them,
both US$_3$ and USe$_3$ are predicted to have 12 Raman-active
vibrational modes \cite{Nouvel87,Zwick79}: 8 A$_g$ and 4 B$_g$ modes
respectively. The symbols A$_g$ and B$_g$ represent for two
different vibration symmetries respectively: vibrations inside and
out of the mirror plane. Specifically, A$_g$ corresponds to the
vibrations where $dy$ is always 0 while B$_g$ corresponds to the
vibrations where $dy$ is not 0.

Our Raman results of US$_3$ and USe$_3$ are obtained by symmetry
analysis based on the vibrational modes calculated within the
GGA+$U$ method, where SOC effects are considered. Figures 6(a) and
6(b) show our determined Raman frequencies for US$_3$ and USe$_3$
respectively, together with the experimental results by G. Nouvel
{\it et al.} \cite{Nouvel87}. One can see that by using the GGA
formalism for both US$_3$ and USe$_3$, the Raman frequencies are
totally confused compared with the experimental results. At another
side, the GGA+$U$ results on the Raman frequencies are clearly
reasonable in two obvious aspects: i) there exist a highest single
Raman frequency for both US$_3$ and USe$_3$, corresponding to the
molecular-like X$_2$ mode vibrations; ii) the highest Raman
frequency of US$_3$ is much larger than that of USe$_3$, indicating
that the S-S interaction is stronger than Se-Se. When comparing all
the Raman frequencies, there seems to be a frequency shift of about
25$\sim$35 cm$^{-1}$ between our GGA+$U$ and the experimental
results.

\section{Conclusion}

By using the GGA and GGA+$U$ methods, we have systematically studied
the electronic, mechanical, and Raman properties of US$_3$ and
USe$_3$, aiming to reveal the underlying electron correlation and
SOC effects. By comparing the calculated lattice constants of US$_3$
and USe$_3$ with corresponding experimental results, and monitoring
their electronic structures with different $U_{\rm eff}$ parameters,
we conclude that a $U_{\rm eff}$ value of 6 eV is needed to get
reasonable lattice constants, and the right insulating properties
for both US$_3$ and USe$_3$. After considering SOC effects, the deep
U-6$p$ band will be split into two separate energy bands in both
US$_3$ and USe$_3$. The SOC effects also changes the energy band
shapes around the Fermi energies and lead to the fact that US$_3$ is
a direct band gap while USe$_3$ is an indirect band gap insulator.
Because of the SOC effects, the energy band gaps of US$_3$ and
USe$_3$ reduce by 0.11 and 0.07 eV respectively. Based on the
obtained antiferromagnetic ground states of US$_3$ and USe$_3$, we
then systematically give out their electrical, mechanical, and Raman
properties. The VBM and CBM of US$_3$ (USe$_3$) are found to be
contributed by 3$p$-5$f$ (4$p$-5$f$) hybrid and 5$f$ (5$f$)
electronic states respectively. The elastic constants of US$_3$ and
USe$_3$ are found to satisfy the stability criteria, and their
Poisson's ratios are calculated to be 0.19 and 0.22 respectively.
For the Raman properties, we theoretically repeat the two important
experimental observations that the S-S or Se-Se dimer vibrations
have the largest frequencies in US$_3$ and USe$_3$, and the
frequency of the S-S dimer vibration is larger than that of the
Se-Se dimer vibration.

\section{Acknowledgement}

This work was supported by the National Natural Science Foundation
of China under Grants No. 10904004 and No. 90921003 and Foundations
for Development of Science and Technology of China Academy of
Engineering Physics under Grants No. 2011B0301060, No. 2011A0301016,
and No. 2008A0301013.

\clearpage
\begin{table}
\caption{Electronic energies per unit cell of US$_3$ and USe$_3$ in
the nonmagnetic (NM), ferromagnetic (FM), and antioferromagnetic
(AFM) states by using different calculation methods. All values are
in units of eV.}
\begin{tabular}{ l   c   c   c   c   c   c }
  \hline
  \hline
  \multirow{2}*{Methods} & \multicolumn{3}{c}{US$_3$} & \multicolumn{3}{c}{USe$_3$} \\
  \cline{2-4}\cline{5-7}
   & NM & FM & AFM & NM & FM & AFM  \\
  \hline
  GGA                          & -55.33 & -55.55 & -55.47 & -50.48 & -50.87 & -50.79 \\%
  GGA+$U$ ($U_{\rm eff}$=1 eV) & -53.30 & -53.82 & -53.78 & -48.46 & -49.30 & -49.31 \\%
  GGA+$U$ ($U_{\rm eff}$=2 eV) & -51.34 & -52.50 & -52.24 & -46.57 & -48.14 & -48.17 \\%
  GGA+$U$ ($U_{\rm eff}$=3 eV) & -49.68 & -51.53 & -51.53 & -45.07 & -47.72 & -47.72 \\%
  GGA+$U$ ($U_{\rm eff}$=4 eV) & -48.42 & -50.85 & -50.85 & -44.50 & -47.26 & -47.35 \\%
  GGA+$U$ ($U_{\rm eff}$=5 eV) & -46.55 & -50.74 & -50.74 & -43.95 & -46.79 & -47.26 \\%
  GGA+$U$ ($U_{\rm eff}$=6 eV) & -48.39 & -51.68 & -51.69 & -43.48 & -46.52 & -46.85 \\%
  \hline
  \hline
\end{tabular}\label{energy} \\
\end{table}
\clearpage
\begin{table}
\caption{Lattice constants of US$_3$ and USe$_3$ obtained by using
different methods. All values are in units of \AA.}
\begin{tabular}{ l   c   c   c   c   c   c }
  \hline
  \hline
  \multirow{2}*{Methods} & \multicolumn{3}{c}{US$_3$} & \multicolumn{3}{c}{USe$_3$} \\
  \cline{2-4}\cline{5-7}
   & $a$ & $b$ & $c$ & $a$ & $b$ & $c$  \\
  \hline
  GGA                          & 5.11 & 3.83 & 9.25 & 5.50 & 3.99 & 10.22  \\
  GGA+$U$ ($U_{\rm eff}$=1 eV) & 5.12 & 3.82 & 9.26 & 5.52 & 3.95 & 10.30  \\
  GGA+$U$ ($U_{\rm eff}$=2 eV) & 5.13 & 3.82 & 9.28 & 5.52 & 3.94 & 10.42  \\
  GGA+$U$ ($U_{\rm eff}$=3 eV) & 5.09 & 3.74 & 9.61 & 5.50 & 3.87 & 10.98  \\
  GGA+$U$ ($U_{\rm eff}$=4 eV) & 5.09 & 3.75 & 9.65 & 5.71 & 4.02 & 10.90  \\
  GGA+$U$ ($U_{\rm eff}$=5 eV) & 5.06 & 3.82 & 9.61 & 5.73 & 4.04 & 10.52  \\
  GGA+$U$ ($U_{\rm eff}$=6 eV) & 5.44 & 3.92 & 9.76 & 5.73 & 4.04 & 10.60  \\
  Exp.$^{a,b}$ & 5.37 & 3.96 & 9.94 & 5.65 & 4.06 & 10.47  \\
  \hline
  \hline
\end{tabular}\label{lattice} \\
\begin{flushleft}
$^a$Reference \onlinecite{Gronvold68} \\
$^b$Reference \onlinecite{Salem84}
\end{flushleft}%
\end{table}
\clearpage
\begin{table}
\caption{Elastic constants of US$_3$ and USe$_3$ obtained in GGA and
GGA+$U$+SOC calculations. The $U_{\rm eff}$ value is chosen to be 6
eV in both GGA+$U$+SOC calculations. All data are in units of GPa.}
\begin{tabular}{ l   c   c   c   c   c   c   c   c   c   c   c   c   c }
  \hline
  \hline
  & $C_{11}$ & $C_{22}$ & $C_{33}$ & $C_{44}$ & $C_{55}$ & $C_{66}$ & $C_{12}$ & $C_{13}$ & $C_{23}$ & $C_{45}$ & $C_{16}$ & $C_{26}$ & $C_{36}$ \\
  \hline
  US$_3$ (GGA)          & -654 &  -80 & -4132 & 452 & -332 & -273 & -923 & -2953 & -1766 &  32 &  -511 & -640 & -522  \\
  US$_3$ (GGA+$U$+SOC)  &  963 & 1055 &   437 & 313 &  299 &  169 &  196 &   132 &   181 &  -6 &  -102 &  -55 &  -74  \\
  USe$_3$ (GGA)         &  946 &  583 &  -107 & 363 &  309 &    7 &  -37 &  -349 &   -65 &   3 &   171 &   24 &  198  \\
  USe$_3$ (GGA+$U$+SOC) &  951 &  938 &   496 & 257 &  244 &  267 &  195 &   194 &   216 & -13 &   -19 &  -20 &  -33  \\
  \hline
  \hline
\end{tabular}\label{elastic}
\end{table}
\clearpage
\begin{table}
\caption{Mechanical properties of US$_3$ and USe$_3$ including the
bulk and shear moduli, their Voigt and Reuss bounds, and the
Poisson's ratio calculated by using the GGA+$U$ method. The $U_{\rm
eff}$ value is chosen to be 6 eV. All moduli data are in units of
GPa.}
\begin{tabular}{ l   c   c   c   c   c   c   c }
  \hline
  \hline
  & $B_{\rm V}$ & $B_{\rm R}$ & $G_{\rm V}$ & $G_{\rm R}$ & $B$ & $G$ & $\nu$  \\
  \hline
  US$_3$  & 386 & 299 & 286 & 245 & 342 & 266 & 0.19  \\
  USe$_3$ & 400 & 365 & 272 & 257 & 382 & 265 & 0.22  \\
  \hline
  \hline
\end{tabular}\label{mechanics}
\end{table}
\clearpage

\noindent\textbf{List of captions} \\

\noindent\textbf{Fig.1}~~~ (Color online). (a) The atomic structures
of MX$_3$, where blue and dark green balls representing for metal
(M) and chalcogen (X) atoms respectively. The monoclinic lattices
are depicted by the dashed lines. (b) Depiction of the
Brillouin Zone for the monoclinic MX$_3$ lattice. \\

\noindent\textbf{Fig.2}~~~ (Color online). The local density of
states (LDOS) for the S1, S3, and U atoms in US$_3$ (a) and Se1,
Se2, and U atoms in USe$_3$ (b) by using the GGA and GGA+$U$ methods
with $U_{\rm eff}$ ranging from 1 to 6 eV. The S1 (Se1) and S3 (Se3)
atoms correspond to X1 and X3 atoms depicted in Fig. 1(a). The Fermi
energies are denoted by dashed lines. \\

\noindent\textbf{Fig.3}~~~ (Color online). (a)-(d) The projected
density of states (PDOS) for the 3$p$ electronic states of S1 and S3
atoms, and 5$f$, 6$p$ electronic states of the U atom in US$_3$.
(e)-(h) The PDOS for the 4$p$ electronic states of Se1 and Se3
atoms, and 5$f$, 6$p$ electronic states of the U atom in USe$_3$.
The S1 (Se1) and S3 (Se3) atoms correspond to X1 and X3 atoms
depicted in Fig. 1(a). The Fermi energies are denoted by dashed
lines. The values of $U_{\rm eff}$ are chosen to be 6 eV in all
GGA+$U$ and GGA+$U$+SOC calculations. \\

\noindent\textbf{Fig.4}~~~ (Color online). The projected density of
states for the uranium 5$f$ electronic states in US$_3$ (a) and
USe$_3$ (b), by using the GGA method. The Fermi energies are set to
be zero and denoted by dashed lines. The results before and after
considering the spin-orbit coupling effects are shown in red solid
and blue dotted lines respectively. \\

\noindent\textbf{Fig.5}~~~ (Color online). The electronic energy
band structures for US$_3$ (a) and USe$_3$ (b) by using the GGA+$U$
method. The values of $U_{\rm eff}$ are chosen to be 6 eV. The Fermi
energy and different $k$-points are denoted by the dashed lines. The
results before and after considering the spin-orbit coupling effects
are shown in solid and dotted lines respectively. \\

\noindent\textbf{Fig.6}~~~ (Color online). The calculated
frequencies at the $\Gamma$ point for Raman-active vibrational modes
of US$_3$ (a) and USe$_3$ (b) by using the GGA and GGA+$U$ methods,
together with the experimental results listed in Ref.
\onlinecite{Nouvel87}. The values of $U_{\rm eff}$ are chosen to be
6 eV. The symmetry for each vibration mode are also presented. \\

\clearpage

\begin{figure}
\includegraphics[width=1.0\textwidth]{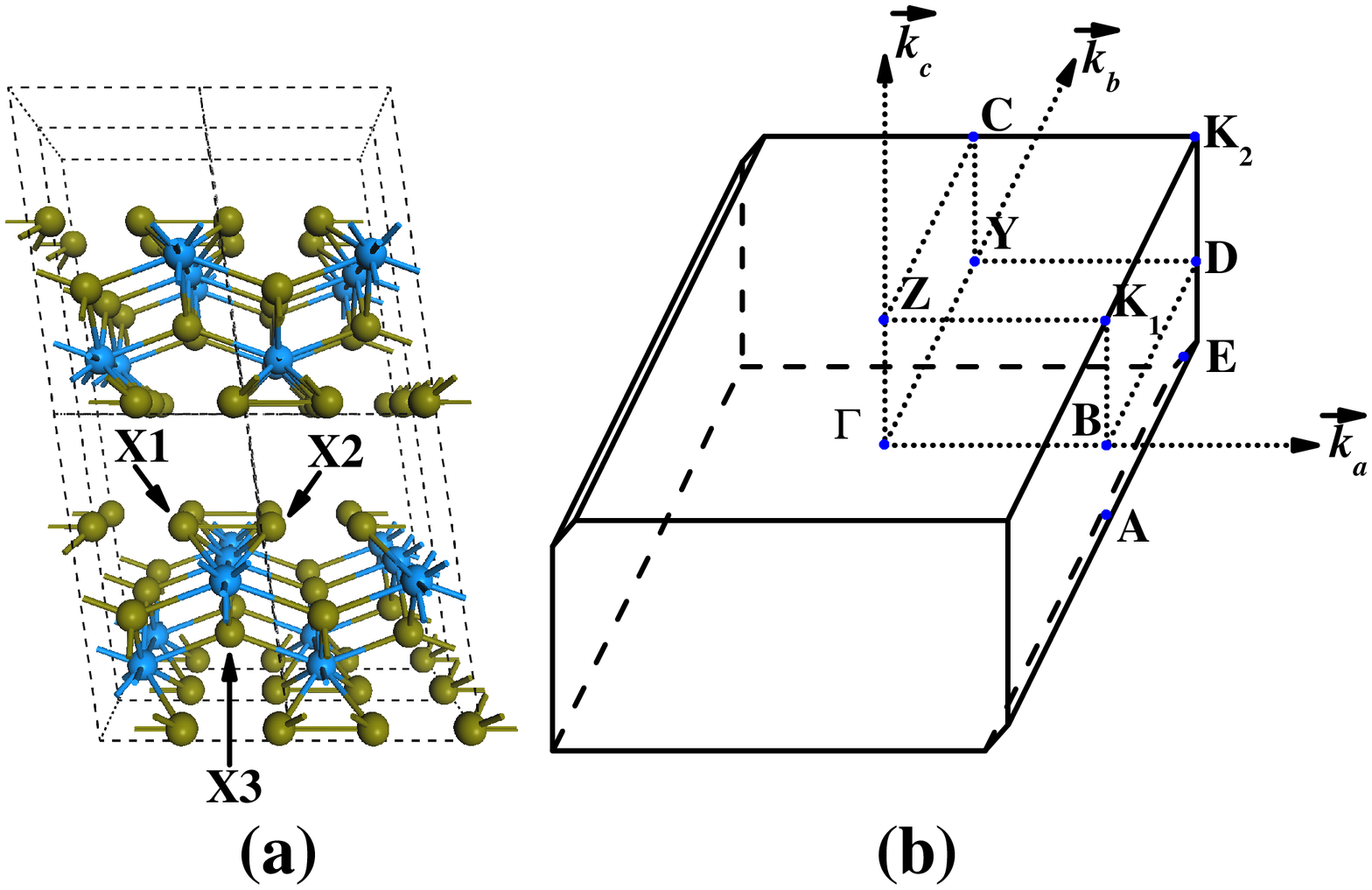}
\caption{\label{fig:fig1}}
\end{figure}
\clearpage
\begin{figure}
\includegraphics[width=0.3\textwidth]{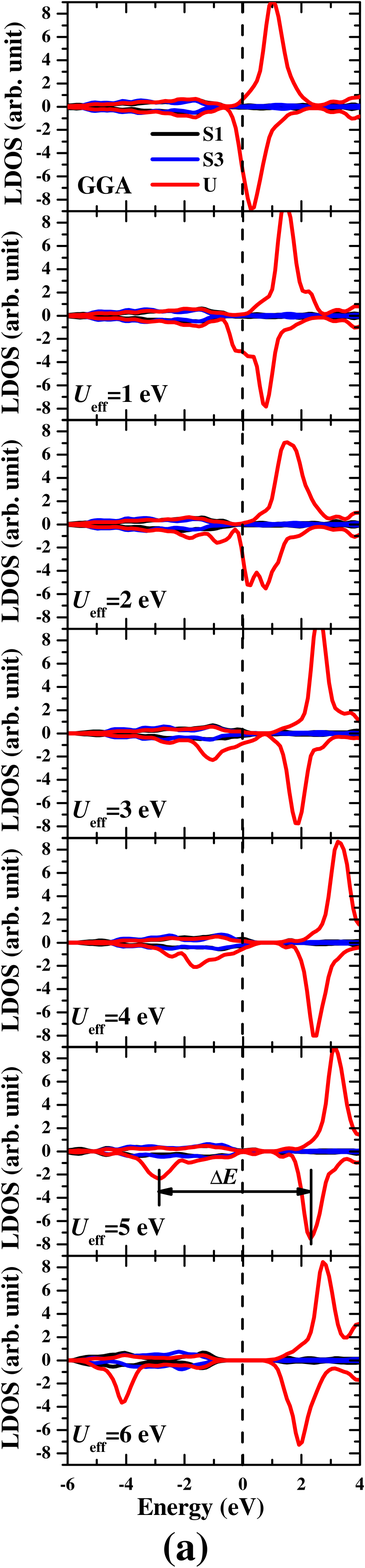}
\includegraphics[width=0.3\textwidth]{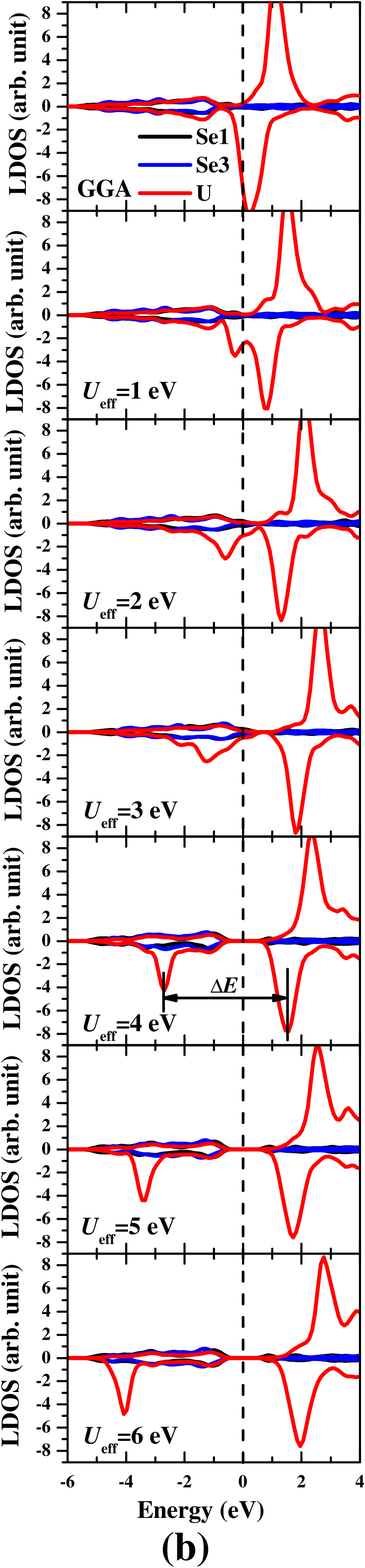}
\caption{\label{fig:fig2}}
\end{figure}
\clearpage
\begin{figure}
\includegraphics[width=0.4\textwidth]{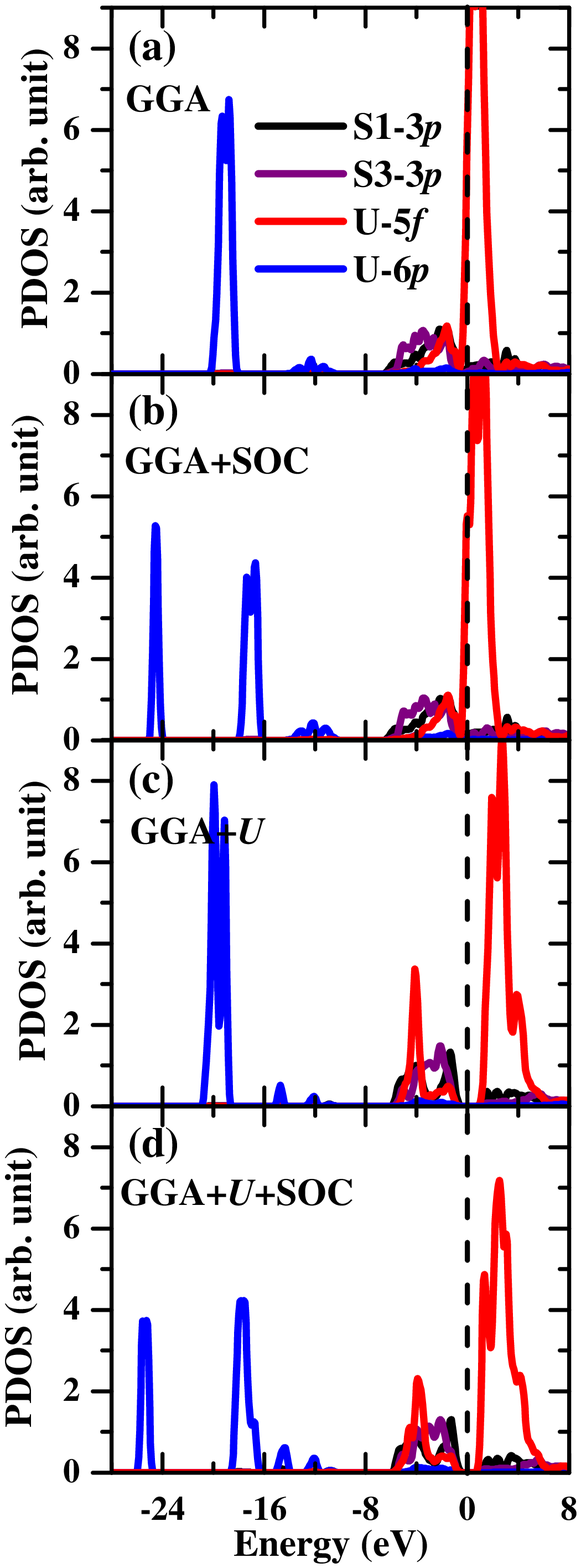}
\includegraphics[width=0.4\textwidth]{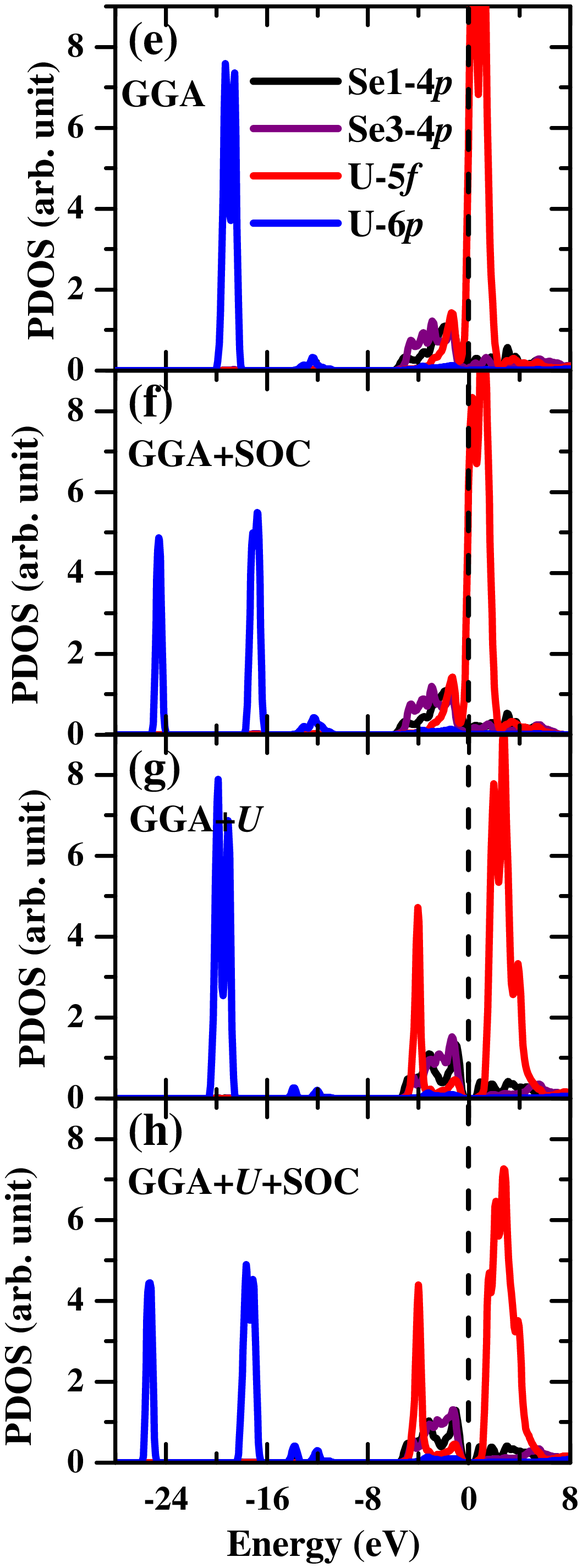}
\caption{\label{fig:fig3}}
\end{figure}
\clearpage
\begin{figure}
\includegraphics[width=1.0\textwidth]{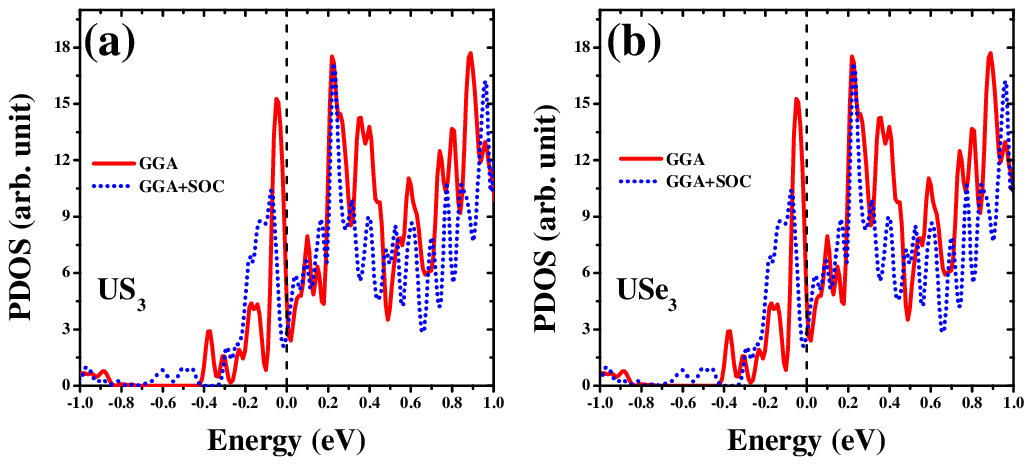}
\caption{\label{fig:fig4}}
\end{figure}
\clearpage
\begin{figure}
\includegraphics[width=1.0\textwidth]{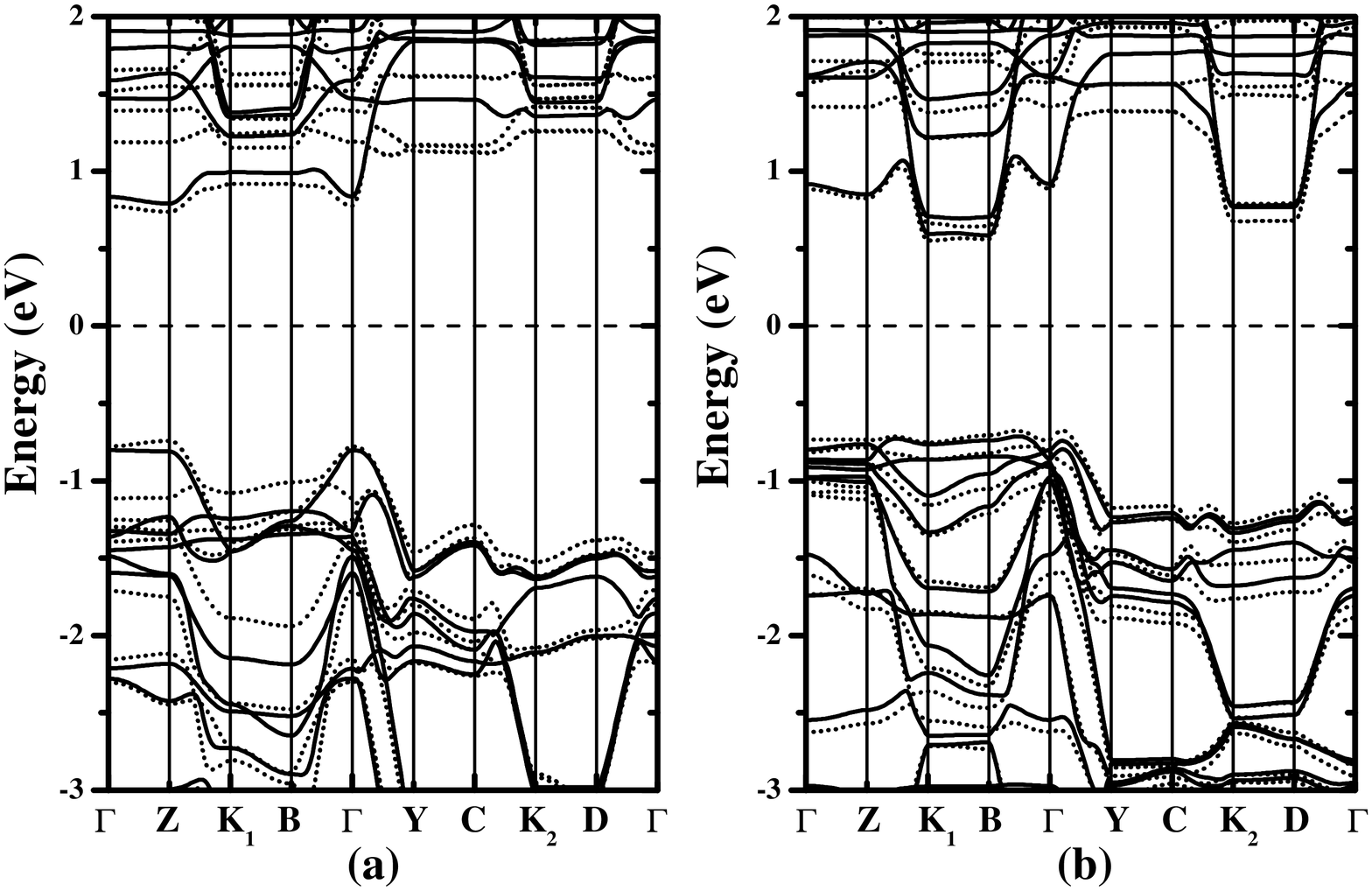}
\caption{\label{fig:fig5}}
\end{figure}
\clearpage
\begin{figure}
\includegraphics[width=1.0\textwidth]{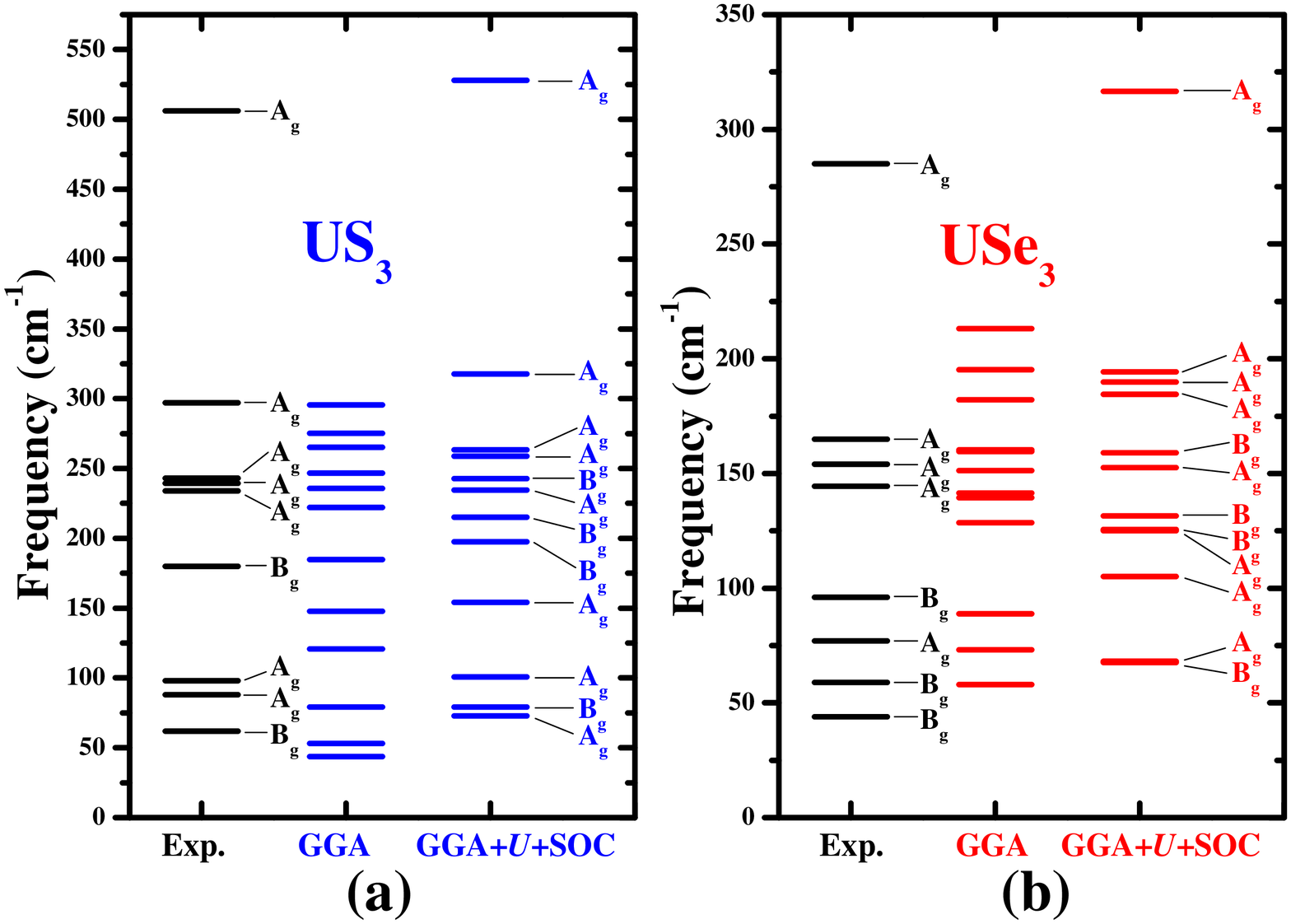}
\caption{\label{fig:fig6}}
\end{figure}
%
\end{document}